\documentclass[12pt,preprint]{aastex}
\begin{document}

\title{Adaptive Optics Nulling Interferometric Constraints on the Mid-Infrared
Exozodiacal Dust Emission around Vega$^{1}$}
\footnotetext[1]{The results presented here made use of the of Multiple Mirror
Telescope (MMT) Observatory, a jointly operated facility of the University of
Arizona and the Smithsonian Institution.}
\author{Wilson M. Liu$^{2}$,
Philip M. Hinz$^{2}$,
William F. Hoffmann$^{2}$,
Guido Brusa$^{2,3}$,
Francois Wildi$^{2,4}$,
Doug Miller$^{2}$,
Michael Lloyd-Hart$^{2}$,
Matthew A. Kenworthy$^{2}$,
Patrick C. McGuire$^{2,5}$,
and J. R. P. Angel$^{2}$}

\altaffiltext{2}{Steward Observatory, University of
Arizona, 933 N. Cherry Ave., Tucson, AZ, USA 85721, email: wliu@as.arizona.edu}
\altaffiltext{3}{Osservatorio Astrofisico di Arcetri, Florence, Italy}
\altaffiltext{4}{EIVD, West Switzerland University of Applied Sciences, Yverdon,
Switzerland}
\altaffiltext{5}{Centro de Astrobiologia (INTA/CSIC), Madrid, Spain}

\begin{abstract}
We present the results of mid-infrared nulling interferometric observations of
the main-sequence star $\alpha $ Lyr (Vega) using the 6.5 m MMT with its
adaptive secondary mirror.  From the observations at 10.6 $\mu$m,
we find that there is no resolved emission from the circumstellar
environment (at separations greater than 0.8 AU) above 2.1\% (3 $\sigma$
limit) of the level of the stellar photospheric emission.  Thus, we are
able to place an upper limit on the density of dust in the inner system of
650 times that of our own solar system's zodiacal cloud. This limit is roughly
2.8 times better than those determined with photometric excess
observations such as those by IRAS.  Comparison with
far-infrared observations by IRAS shows that the density of warm dust in
the inner system ($< 30$ AU) is significantly lower than cold dust at larger
separations.  We consider two scenarios for grain removal, the sublimation
of ice grains and the presence of a planetary mass "sweeper."  We find that
if sublimation of ice grains is the only removal process, a large fraction ($>
80$ \%) of the material in the outer system is ice.
\end{abstract}

\section{Introduction}
Circumstellar material around main-sequence stars was first discovered in
1983 by the Infrared Astronomical Satellite (IRAS), when observations of
$\alpha$ Lyr (Vega) showed far-infrared (FIR) emission in excess of that
expected from the stellar photosphere \citep{aumann}.  These detections were
made at 60 and 100 $\mu$m and were followed by similar discoveries around other
stars, including $\beta$ Pic and $\alpha$ PsA \citep{gillett}.  Since then,
many stars have been found to have excess FIR emission, and this emission has
been thought to be associated with thermal emission from cold circumstellar
debris, similar in nature to the Kuiper disk surrounding our solar system.

Though several cases of this so-called "Vega Phenomenon" have been confirmed
at FIR wavelengths, to date there have been no resolved detections of
mid-infrared (MIR) emission surrounding main-sequence stars older than a few
tens of Myr.  Such
a detection would be indicative of warm ("room-temperature") dust close to
the star, analogous to the zodiacal dust in our own solar system.  Any material
emitting at 10 $\mu$m would be in the "habitable zone" of a system where liquid
water could exist.  The presence of dust would also necessitate the presence of
planetesimals which would, through collisions, regenerate the dust that is
normally depleted on fast timescales due to Poynting-Robertson drag and/or
radiation pressure blow-out.  Resolved MIR disks have been detected
around the main-sequence A-type stars $\beta$ Pic \citep{pantin,wein03} and
HR 4796A \citep{jaya98,jura98}, though
HR 4796A is thought to be very young, with an age less than 10 Myr and $\beta$ Pic
is estimated to be less than 30 Myr old \citep{barrado}.  Detection
of a MIR disk around an older star like Vega (350 Myr old; \citet{lach99})
proves to be more difficult, since stars of this age are thought to have shed
their natal circumstellar disks.  Any new dust generated in the manner described
above would have a total flux several orders of magnitude below that of the
host star.

In order to mitigate the difficulty of observing such a faint source in the
presence of a bright star, our observations of Vega make use of nulling
interferometry and the adaptive secondary mirror for the MMT.
Nulling interferometry is a relatively new
technique to detect resolved faint structure in the presence of a much brighter
unresolved point source.  The BracewelL Infrared Nulling Cryostat (BLINC)
uses two parts of the MMT's 6.5 m primary mirror to create an
interferometer with two elliptical 4.8 x 2.5 m subapertures and a baseline of
4 m.  These two subapertures are overlapped in the pupil plane with
an appropriate path difference between the beams to destructively interfere
the central point source in the image plane. The effect of this optical
arrangement is to create a sinusoidal transmission pattern for the object on the
sky, with destructive interference on the point source (star). Light from
half an interference fringe width away constructively interferes, enhancing the
flux. Thus the technique can detect material as close to the star
as one-quarter of the fringe spacing where the light is neither suppressed
nor enhanced.  This corresponds to 0.12 arcseconds for the configuration used
on the MMT (for Vega, this is a projected separation of less than one AU).
Nulling interferometry is unique in that it can be
used to determine the relative contributions of the star and circumstellar
material to the total flux, independent of models of the stellar photosphere
and nearby environment.  Full details of the BLINC instrument and its
implementation can be found in \citet{hinz_phd}.
The MMT's adaptive secondary mirror provides correction
of atmospheric wavefront aberrations in the incoming light, allowing
destructive interference to be precisely tuned for the deepest possible
suppression of starlight.

In this Letter, we present results from 10.6 $\mu$m nulling interferometric
observations of the main-sequence star $\alpha$ Lyr (Vega), which are part of
a larger systematic survey of solar neighborhood main-sequence stars.  We
discuss constraints of the distribution, density, and composition of the dust
in the Vega system and compare our results with previously published observations
of Vega at different wavelengths.

\section{Observations and Data Reduction} \label{sec-obs}
Observations were made in May 2003 at the MMT 6.5 m telescope on Mt. Hopkins,
Arizona.  The observations made use of the world's first adaptive optics (AO)
secondary mirror.  Since the deformable mirror is the telescope's
secondary mirror, there is no need for an intermediate set of reimaging and
correcting optics as in traditional AO systems.  Thus, the light from the
secondary is fed directly into the science camera, optimizing throughput and
decreasing the background emissivity in the MIR by avoiding the use of extra
warm optics.  Wavefront sensing is accomplished in a separate assembly using
visible light diverted from the telescope beam.  The wavefront sensor is a
Shack-Hartmann sensor with a EEV CCD39a detector, operating at a frame
rate of 550 Hz.  For further details regarding the MMT AO system, we refer the
reader to \citet[and references therein]{mmtao}.

The main advantage in using AO with nulling interferometry
is the stabilization of the wavefront of light, allowing
us to precisely adjust the path difference between the two beams of the
interferometer, and thus obtain the deepest possible null.  However, in the
case of these observations, the suppression of starlight was limited during
the run by a slight mechanical vibration (measured during subsequent observations
to be about 20 milliarcseconds at a
frequency of 20 Hz) in the telescope, which caused a wavefront aberration and
therefore a slight phase error between the two beams of the interferometer.
This caused the instrumental nulls to vary slightly and limited the suppression
to the levels described below (about 3\% residual light), where theoretically
a deeper null (several tenths of a percent) would be possible.

Nulling observations were taken with a broadband N filter  (8.1- 13.1 $\mu$m).
For the science object, Vega, seven sets of twenty frames were taken with the star
destructively interfered, for a total of 140 frames,
each frame with an integration time of 3 s.  Observations of Vega in
constructive interference were taken between the destructive sets, with the
same integration times.  All frames were sky subtracted using off-source
frames taken in between each set of observations of Vega.  The "instrumental
null" for each destructive frame was calculated by simply taking ratio of the
flux of the nulled (destructively interfered) image to the flux of the
constructively interfered image (i.e., instrumental null = nulled flux / full
flux), expressed as a percent. Each destructively interfered set of
frames was examined for the frame with the best instrumental null.  In order
to calibrate these values, observations of two point-source (unresolved)
standard stars, $\alpha $ Her and $\gamma $ Dra, were taken before and after
the observations of Vega.

The best instrumental nulls for the 7 destructively interfered sets of images
are shown in Table \ref{tab:obs}.  Each set of frames was taken
at a different orientation of the interferometer baseline relative to the
sky, which allows us to probe for evidence of an inclined disk structure, if
resolved emission is detected (see \citet{liu03} for an explanation of this
technique).  The standard deviation in the derived null for the sets of
frames is 0.7\%, resulting in an average value of the null of 3.7\% $\pm$ 0.6\%.
The nulls for the individual point-source calibrators were 3.6\% $\pm$ 0.5\%
from 3 sets of observations of $\alpha$ Her, and 3.4\% $\pm$ 0.4\% from 7 sets
of observations of $\gamma $ Dra.  This results in a combined average null on
the point-source standard stars of 3.5\% $\pm$ 0.4\%.
"Source nulls" were calculated for Vega by subtracting the null obtained
for the standard stars from the instrumental
null achieved for Vega (i.e, Source Null = Instr. Null - Standard Null).  This
value represents the flux of resolved emission around a star, as a percentage
of the full flux of the star when constructively interfered.  A non-zero source
null means that resolved emission has been detected.

\section{Results and Discussion}
The source null derived for Vega is 0.2\% $\pm$ 0.7\% (1 $\sigma$ error),
consistent with zero,
which indicates that we are not detecting resolved emission at our current
levels of sensitivity and spatial resolution \footnote{In contrast, for an
example of a positive detection of circumstellar material, see \citet{liu03}}.
This allows us to place
constraints on the distribution and amount of exozodiacal dust surrounding
Vega.  We are confident (3 $\sigma$) that there is no resolved emission at 10.6
$\mu$m around Vega above the 2.1\% level (0.9 Jy) outside of 0.8 AU from the
star.  In order to compare this limit with the zodiacal dust density in our
own solar system, we use the zodiacal dust model of \citet{kelsall}.  The Kelsall
model would result in a nulled flux of 1.4 mJy (or 0.0033\% of Vega's flux) if
placed at the distance of Vega.  Scaling up this solar  model to our 3 $\sigma$ limit
for Vega corresponds to a dust density limit of about 650 times our solar system's
zodiacal dust.  Additionally, we find that the null does not vary significantly with
observations at different rotations of the interferometer baseline (over a range
of about 90 degrees), indicating that there is no evidence of an inclined disk-like
structure.

A further analysis can be made by comparing the observed limits to an estimate
of the expected flux of our own zodiacal dust \emph{at the age of Vega (350 Myr)}.
For a collisionally replenished disk with a dust removal timescale much shorter
than the lifetime of the system, we expect $f_{d} \sim t^{-2}$, where $f_{d}$ is
the dust flux as a fraction of stellar flux \citep{spang}.  Using this relation we
find that the transmitted signal from our zodiacal dust at the age of Vega (350 Myr)
would be $\approx 270$ mJy.  Our limit of 0.9 Jy results in a limit on warm dust
in the Vega system of about 3 times our own zodiacal dust, after accounting for
dust evolution.

Previous studies of Vega in the FIR have found a significant amount of cold
(50 - 125K) debris in the system at large separations (several tens of AU and
greater) \citep{bp93}. From Poynting-Robertson drag one would expect that this
material would migrate inward and populate inner regions with material as well,
which could be detected at MIR wavelengths.  If one assumes that the excess
flux at 25 $\mu$m (1.08 Jy, \citet{bp93}) is
due to thermal emission from cold debris, takes the temperature of grains as a
function of distance from the star as T $\sim r^{-0.5}$ \citep{bp93},
and makes the conservative assumption
that the optical depth profile of the circumstellar material is constant with
radius \footnote{For reference, the solar system's radial optical depth profile is
$\sim r^{-0.4}$ in the inner system (r $<$ 30 AU) and $\sim r^{-4}$ for
30 AU $>$ r $>$ 100; \citet{back97}.  For our analyses, we assume that the
grains are large enough to be considered blackbodies.},
one would expect a flux of 11 Jy from blackbody grains at 10 $\mu$m.  This flux is
calculated assuming thermal blackbody emission from the grains by integrating the
product of the Planck function and optical depth over the spatial extent of the
zodiacal dust \footnote{We set the inner radius of the disk at the dust
sublimation radius of 0.25 AU, corresponding to a temperature of 1500 K,
and the outer radius of the disk
to the distance at which we expect the thermal emission at a given wavelength
to peak given the T vs. r relation in \citet{bp93}. The dust is assumed to be
optically thin.}.  When this signal is
observed through the transmission pattern of the interferometer, we estimate
the final signal to be over 5 Jy.  Using the 60
$\mu$m excess (7.75 Jy, \citet{bp93}), with the same assumptions as above, the
calculation yields an even greater detected excess of 470 Jy at 10 $\mu$m.  These
expected 10 $\mu$m fluxes would have been easily detectable with our observations.
However, we do not find this large excess, which indicates that
the inner region of the Vega system is relatively clear of material compared
to the outer region. This result is consistent with previous conclusions
\citep{lba00,bp93} though our observations are able to better constrain the
upper limit of dust density in the Vega system by a factor of $\approx $ 2.8 times
compared to IRAS observations, which provided an upper limit for warm dust
density of 1800 times our solar system \citep{hinz_phd,aumann}.

The lack of substantial 10 $\mu$m emission surrounding Vega can be physically
interpreted in different ways.  One may draw a comparison to the HR 4796A
system which was observed by \citet{jura98} to have a similar lack of warm
material in the inner system.  For the case of HR 4796A, they
suggest two possible scenarios for the lack of warm dust in that system: the
existence of a companion clearing out material, or the destruction of ice particles
by stellar radiation.  Here we consider the same explanations for the absence
of material in the inner Vega system.  For the latter scenario, we would expect
water ice to sublimate at temperatures above 110 K \citep{poll94}, and in
fact we may already see evidence of this effect in the 25 $\mu$m emission which
probes temperatures near the sublimation temperature, and shows a smaller than
expected excess compared to longer wavelengths.  If the lack of material in the inner
system is due to sublimation of ice grains, we can constrain the composition
of the outer debris disk. To estimate this effect we make the simplifying assumption
that each grain is composed of either icy material that sublimates at 110 K and
thus is totally destroyed inward of about 45 AU, or silicates/metals grains
which remain unaffected.  Comparing our derived upper limit on dust
flux at 10 $\mu$m (0.9 Jy) and the expected 10 $\mu$m flux using the 60 $\mu$m
excess calculated above (470 Jy), and assuming the density decrease in the
inner system is due only to the sublimation of ice grains, this would require
that the outer disk is comprised of 99.8\% icy material, assuming optically
thin material.  Using the 25 $\mu$m excess to estimate the density contrast
compared to the inner system, we find that the composition must be about
80\% water ice, again if ice sublimation is the only cause of grain removal.
For comparison, the large  Kuiper Belt Objects Pluto and Charon
have densities of roughly 2 g/cm$^{3}$ \citep{luu}, indicating higher
fractions of silicate material in our own outer solar system than that
of Vega, as determined by this study. This either suggests a significant
difference in composition, or points to another explanation for removal of
the cold dust as it spirals in.

Another explanation for the lack of dust in the
inner system is the presence of a sweeper companion.  Previous observations in the
millimeter and submillimeter \citep{wilner,koerner,holland} have detected dust
in the Vega system at separations between $8\arcsec$ and $14 \arcsec$ from the star
(projected separations of 60 to 110 AU).  The morphology of the dust is suggested
to be the result of a planetary perturber.  For example \citet{wilner} suggest
a planetary companion of 3 $M_{Jup}$ at a separation of about 50 AU.  It is
conceivable that such a companion could be responsible for the contrast in
density of circumstellar material between the outer and inner system found
in this and previous studies.  However if one assumes that the temperature
of grains follows the relation $T_{g} \sim r^{-0.5}$ \citep[eqn. 3]{bp93},
the drop off in mid-infrared excess between the 25 $\mu$m IRAS detection and the
10 $\mu$m observations in this study indicate a significant density decrease
between 10 and 40 AU, suggesting that a planetary companion may be located at a closer
separation than suggested by the millimeter observations.  Recent near-infrared
adaptive optics observations of Vega by Keck \citep{mac03} and the Palomar
5 m \citep{met03} have attempted to detect planetary mass companions.  These
studies found no evidence for a massive ($>$ several $M_{Jup}$) planetary
companion.  However, the studies note that they do not probe masses for companions
as low as those suggested by the millimeter observations.

Finally, we compare our results to those of \citet{ciardi}, who find near-infrared
emission consistent with a circumstellar debris disk within 4 AU emitting at 3\% - 6\%
of the stellar flux.  If we take this to be the case, and assume that the optical
depth of material drops off as r$^{-0.4}$ \citep{back97} out to the 10 $\mu$m
emitting region, we would
expect a signal in the range of 1.5 - 3 Jy at 10 $\mu$m, which would have
been detected
by our observations.  We do not find this to be the case, which suggests
that if a near-infrared disk is present, there is a steeper drop off in the optical
depth of dust than the r$^{-0.4}$ assumed here.

\section{Future Directions}
We are currently in the process of refining the nulling technology in order to
achieve better suppression of starlight.  Technical improvements include the
addition of an internal servo loop in BLINC, which will actively monitor and
correct the phase difference between the two beams of the interferometer in
order to maintain the deepest possible null.  Stabilizing the phase error
will allow us to suppress the starlight by a factor of $\sim 1000$. 
Scientifically, this improvement in the suppression will allow us to detect
levels of zodiacal emission many times fainter than currently possible, in the
50-100 zody level.  Using the refined nulling observations, we plan a survey of
nearby main-sequence A-type stars to search for evidence of exozodiacal dust.
These observations also lay the groundwork for a planned survey with the Large
Binocular Telescope Interferometer, which will carry out nulling searches for
exozodiacal dust at sensitivities approaching solar level.

\section{Acknowledgements}
W.L. was supported under a Michelson Graduate Fellowship.  The authors thank
the staff at the MMT, especially the telescope operators, M. Alegria, J. McAfee,
and A. Milone, for excellent support.  We thank B. Duffy for technical support with
the BLINC/MIRAC instrument.  W.L. is grateful to A. Marble for valuable assistance
during analysis of the data.  The authors also thank the referee, Dr. Gerard
van Belle, for helpful
comments in the revision of the paper. BLINC was developed under a grant from
NASA/JPL, and MIRAC is supported by the NSF and SAO.  The MMT AO system was
developed with support from AFOSR.

\clearpage
\begin{deluxetable}{ccc}
\tablecaption{Summary of Observations of Vega (see \S \ref{sec-obs} for
explanation) \label{tab:obs}}
\tablewidth{0pt}
\tablehead{
\colhead{Set} &
\colhead{PA (\degr)} &
\colhead{Instr. Null (\%)}
}
\startdata
1 & 72 & 2.7\\
2 & 73 & 2.9\\
3 & 53 & 4.4\\
4 & 36 & 4.5\\
5 & 17 & 3.9\\
6 & -21 & 4.0\\
7 & -22 & 3.6\\
\enddata
\end{deluxetable}

\clearpage

\begin{figure}
\caption{Constructively and destructively interfered images of $\alpha \ $Lyr.  The
instrumental null is $3 \pm 1$\%.  The elliptical shape of the star is due to the
ellipical shape of the subapertures in the interferometer.  The additional
structure seen in the constructive image (above and to the right of the central
image) are part of the star's Airy pattern.}
\label{fig:aonull}
\end{figure}

\end{document}